\documentclass[keys,showpacs,preprintnumbers,amsmath,amssymb]{revtex4}

\usepackage{graphicx}
\usepackage{dcolumn}
\usepackage{bm}
\begin{document}
\preprint{pre-print number}
\title{Evolutionary dynamics and diversity in populations}
\author{Juan G. Diaz Ochoa}
\email{diazochoa@itp.uni-bremen.de}
\affiliation{\mbox{Fachbereich 1, University of Bremen, Otto Hahn Allee, D-28359 Bremen}}
\begin{abstract}
 The dynamics of populations is rich, taking into account that both, the individual's actions and the population's fitness are coupled. The way in which an individual chooses a strategy depends off course on the interaction with other individuals and the relation between selection and mutation within the population. The present model considers individuals with {\it memory}. This memory is represented by a device where information of past actions is stored as bits in a 1D Ising chain. The selection of a new individual action depends on the individual's memory. If the selection of a strategy does not improve the individual's fitness, a new individual with different memory size replaces it. Both, actions and memory are observables that characterize the population. They can change in time, and both depend on the fitness of the population. This model allows the implementation of {\it learning} parameters as well as an {\it external information source}, acting as an {\it external field} which drives individuals to select one preferred action. In particular we show that the {\it diversity} of the population, measured as a {\it Shannon's diversity index} (equivalent to a neg-entropy), is not only related to the {\it energy consumption and size} of the system, but is also related to the way in which the individuals are influenced by the {\it external field}.
\end{abstract}
\maketitle

\section{Introduction}
The concept of natural selection, proposed by Charles Darwin and Alfred Russell Wallace, is underlyed by a fundamental driving force that explains how the different types of species are in a permanent flux. Several models apply this principle to modelate different kinds of populations using simple concepts, namely mutation (i.e. changes in the genotype and phenotype of the individuals) and selection (according to the fitness differences between the elements of a population). Additionally, a random drift in the whole population must be included in order to account for fluctuations and effects of noise in the population \cite{Traulsen}. In the recent years, some works have been analyzed fundamental mathematical aspects of co-evolutive dynamics in populations assuming frequency dependent selection (A good review is given by Nowak et. al. \cite{Nowak}; an estandart reference is the book of Hofbauer and Sigmund \cite{Sigmund}). However, this concept of evolution does not only concerns only dynamics of populations, but also systems in economy \cite{Berninghaus} and even molecular biology \cite{Istvan}.

General abilities are usually assigned to competing populations of individuals of two different classes $A$ and $B$, in order to model mutation-selection processes. Mathematically, such processes are usually modeled by game theory, where two individuals playing a game increase or decrease their scores according to the actions they adopt \cite{Sigmund}. In the context of biological sciences, the score of the game is equivalent to the fitness of an individual. Furthermore, if the individual plays in a rational way, the probability to obtain a higher fitness increases. Therefore, the selection of a rational action depends on whether selfishness or altruism may improve the final fitness of the individual \cite{Nash}.

Based on this principle, a kind of phase behavior can be defined for infinite populations. If the fitness (defined by a pay-off matrix) is used as a control parameter, then it is possible to control the number of individuals of both classes $A$ and $B$ into the population (See for example \cite{Traulsen_I}, where the interaction matrix is rescaled; Lee et al. \cite{Lee} make an analysis of games with different pay-off functions). However, population dynamics cannot be simply reduced to this phenotypic point of view. Internal characteristics (which can be more or less simple), related with the way in which individuals try to find out in an even more rational way the best action, must also be included.

Within the present work, the phenotype will be related to the action each individual adopts; this phenotype is a dynamical variable that depends on the perception and memory of individuals, which is of variable size. Hence, the selection of a new action is based on the information stored from past moves, i.e. is a rational choice based on the direct information of the game (and indirectly on the score each individual obtains). The individual's action evolves in order to adapt itself to the changes in the fitness. Simultaneously, the size of the storage device evolves depending on the variation of the individual's fitness. From a biological point of view, the memory size is related to the individual's genetic pool, because it influences the strategies that each individual adopts as response to the pressure of the evolution dynamics \cite{Lindgren}. Hence, depending on the actions, the genetic pool can mutate in order to increase the fitness. Thus, individual's identity is changing. 


Several works make an analysis of the effect of learning schemas in the co-evolutive dynamics and equilibrium of games \cite{Lindgren}. C. Hauert et al. \cite{Hauert} also analyze the effect of the pay off matrix in the evolution of strategies in agents with large memories; an alternative schema in economics is proposed by R. Selten and R. Stoecker \cite{Selten}). In these works the storage device is implemented using a linear schema, as a kind of genetic code. Other models have used notions of learning implemented into the dynamics of the whole population and not into the dynamics of the individual itself \cite{Traulsen_II}. Such schemas are not suitable to implement large memories. Additionally, they do not allow the implementation of an analysis of the system under different cognitive scenarios, in particular under the influence of {\it external information sources}. 

The present model combines the notion of an Ising learning schema and game theory \cite{Engel}, joining the stochastic of the population with the stochastic of a primitive perception model. Such a model allows us to implement individuals with larger memories, implying a more complex dynamics of the memory distribution. Furthermore, with this schema it is possible to model primitive learning aspects. We are also able to analyze the effect that external information may have onto the system (similar to an {\it external field}). Such external information source induces a kind of {\it normative imitation} into the individuals' \cite{Lesourne}. The selection of strategies is therefore not only influenced by the dynamics of the population, but also by the dynamics related to the individual learning schema. The actions as well as the storage size co-evolve and are two observables in this model.

The consideration of this dynamical memory as an observable of the population is a way to define diversity in the population accurately. From a biological point of view, the observation of the distribution of actions is a measure of the phenotypic diversity; at the same time the observation of the distribution of memories is a measure of the genetic diversity. The basis of both distributions is a source of energy provided by the environment. In his book published in 1944, Schr\"{o}dinger explains that life maintains order by degradation of free energy and producing high entropy waste \cite{Kleidon, Schrodinger}. However, this argument seems to be not enough to explain why diversity is not uniformly distributed. In the present model the {\it diversity} is measured as a function of the {\it number of microstates with a given memory}. This observable has been overlooked, ignoring its wide-ranging effects on community structure and ecosystem process \cite{Crutsinger}. Given that this distribution depends on the pay-off and the internal energy of the individuals, as well as the external information, we are able to explore the {\it Shannon's distribution index}, equivalent to the negative entropy of the system, as a function of the learning characteristics of the individuals as well as the influence of external information sources (for example genetic material transported by water, wind or insects).

An important question to solve is if the evolutive process takes place at the individual's or at the population's level. If the dynamics of the population is stochastic, then local variations of the processes may introduce qualitative changes in the dynamics \cite{Traulsen_I}. In the present approach individual's information processing introduces an additional element into the system. The whole population affects the fitness of each individual. But, if the individual is able to process information in order to produce an action, then the coevolutive process does not have only origin at the population level, i.e. individual characteristics influence the co-evolution of the population. 

In the next section the fundamental definitions and the mathematical basis of the model are introduced. In the first subsection the mathematical basis of game theory is explained; in the second and third subsections the dynamics of the actions and evolution of the memory size is explained. The last subsection contains the implementation of the model. In section number three are two big parts: one for the analysis of the distribution of actions and a second for the analysis of the distribution of memories. In the last section a discussion of the results is presented.

  



\section{Dynamics for a coevolutive system with memories}

\subsection{Fundamental definitions for the mathematics of population dynamics}
 We describe a population of $N$ individuals. Each individual adopts an individual action $\sigma^{i}$ that influences its fitness. According to this action, each individual can be of two types, either $C$ (for individuals that cooperate) or $D$ (for individuals that defect). Additionally the individuals can observe and store its own as well as the opponent's actions. With this information the agent can decides the new action. The size of the storage device (Memory size) $m_{i}$ is also an individual characteristic and is defined as an observable. 

The action of individual $i$ is represented by a unitary vector ${\bf \sigma^{i}}$, that can be either ${\bf \sigma^{i}}=(1,0)$ or ${\bf \sigma^{i}}=(0,1)$ for the two classes of actions $C$ and $D$ respectively. The pay-off in the system depends on these actions and is given by the following matrix

\begin{equation}
U^{ji}={\bf \sigma^{j}}{\bf F_{p}}{\bf \sigma^{i}},
\label{UTIL_I}
\end{equation}
where $\sigma^{j}$ is the opponent's action, ${\bf F_{p}}$ is the fitness associated to the action of both individuals $i$ and $j$. In the present model only pair interactions are allowed. The fitness matrix is represented by
\begin{equation}
{\bf F} =\left(\begin{array}{cc}
R & Q \\ 
S & P 
\end{array}\right).
\end{equation}
The scenario implemented in this work is for a prisoner's dilemma game, i.e. the matrix has the values $Q>R>P>S$, where $Q$ is for temptation, $R$ for reward, $P$ for punishment and $S$ for sucker. In this model $Q = 5$, $R = 3$, $P = 2$ and $S=0$. Using eq. (\ref{UTIL_I}), the outcome for the individual $i$ can be computed as
\begin{equation}
f^{i} = \sum_{j=1}^{K}U^{ij},
\label{utility}
\end{equation}
where $K$ is the number of neighbours. The utility is a fundamental quantity that should determine the co-evolutionary process.

\subsection{Dynamics of the actions}
\begin{figure}[h!]
\begin{center}
\includegraphics[clip,  width=0.40\textwidth]{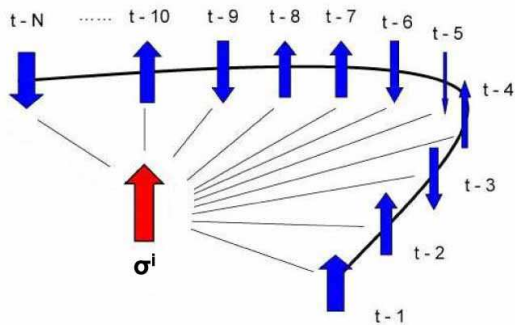}
\caption{This diagram represents the 1D Ising memory. The principal action depends on the influence of the stored actions from the past, modeled as individual spins in the memory of the individual. The new action ${\bf \sigma^{i}}$ is adopted if it minimizes the energy asociated to the chain of spins in the memory.}
\label{figIssing}
\end{center}
\end{figure}

{\it Memory} is defined as an individual device where the information of previous actions of the individual and its neighbours is stored. Using this information a new action can be adopted. There are several techniques that simulate how this storage of information and implementation of new moves takes place; usually such techniques are based on linear methods \cite{Lindgren}. In the present work, the memory works like a field that influences the dynamics of the new action (That means the memory is associated to a kind of internal energy; see Fig. \ref{figIssing}). Using this schema there is a relation between the perception of the individual (observation of the actions in the game) and the perturbations and errors implemented in the game (noise).

An Ising schema is adopted. For simplicity, the perception lattice in the 'mind' of the individual is one dimensional. Each of the actions in the past are bits stored in the memory of the individual and are coupled to the new action. This new action is adopted if the internal energy is minimized. The individual's actions depend off course on the appropriate selection of the internal coupling parameters.  

The Hamilton function for the Ising perception of the individual $i$ depends on the individual's past actions $i$ and its opponents $j$ 
\begin{equation}
\mathcal{H}_{0}^{i}=\sum_{l =1}^{m^{i}_{l}}J_{il}{\bf \sigma^{i}(t)}{\bf \sigma^{l}}(t-l) + \sum_{j =1}^{m^{i}_{j}}J_{ij}{\bf \sigma^{i}(t)}{\bf \sigma^{j}}(t-j),
\end{equation}
where ${\bf \sigma^{l}_{\tau}}(t)$ and ${\bf \sigma^{j}}(t)$ are respectively the past actions of individual $i$ and individual $j$ stored in the memory of individual $i$; $J_{il}$ and $J_{ij}$ are the coupling constant between the new action and the internal stored states of respectively the individual $i$ and its opponent $j$. $m^{i}_{j}$ and $m^{i}_{i}$ are the number of bits that the individual can store from respectively own and opponents past actions. The total storage size is defined as $m^{i}=m^{i}_{j}+m^{i}_{l}$ 


Additionally an external field can also be introduced. In this context, this external field represents the influence that external information may have in the individual's decision. This should induce a normative imitation by means of an external influence \cite{Lesourne}. An external influence is for example a kind of signal trying to orient the actions of the individuals to either of both states $C$ or $D$. The individual's internal energy function adopts the following form 
\begin{equation}
\mathcal{H}^{i}=\sum_{i=1}^{m^{i}_{l}}J_{il}{\bf \sigma^{i}}(t){\bf \sigma^{l}}(t-l)+\sum_{j=1}^{m^{i}_{j}}J_{ij}{\bf \sigma^{i}}(t){\bf \sigma^{j}}(t-j)+ \sum_{k=1}^{m^{i}}H{\bf \sigma^{k}}(t-k),
\end{equation}
 where $H$ is the intensity of the external field. When $H>0$ the individuals have a normative tendency to cooperate. Otherwise, the individuals tend to defect. 

The state ${\bf \rho^{i}}$ is the probability that the action of the individual $i$ is in a given state $c$ is described by a master equation defined on the ensemble of the storage element
\begin{equation}
\frac{d{{\bf \rho^{i}}_{c}}(t)}{dt} = - \sum_{c \neq d}[ W^{i}_{c \rightarrow d}{\bf \rho^{i}}_{c}(t) - W^{i}_{d \rightarrow c}{{\bf \rho^{i}}}_{d}(t)].
\label{Master_I}
\end{equation} 
The transition probabilities in the equation eq. (\ref{Master_I}) can be defined such that it reaches an equilibrium state, implying a balance between the entropic and the energetic coefficients. This balance is reached when the individual optimizes its stored information. Hence, a transition probability of Metropolis form is adopted. The action of the individual $i$ can flip with a transition probability
\begin{equation}
W^{i}_{c \rightarrow d}=\frac{e^{-\beta \Delta \mathcal{H}^{i}}}{T},
\end{equation}
where $\beta=\frac{1}{\Lambda}$, $\Lambda$ the learning parameter, equivalent to the level of noise in the environment \cite{Engel}, $T$ is the period of time and $\Delta \mathcal{H}^{i}$ is the difference between two internal states. The mechanism for the definition of new actions is based on the perception of the information of the 'eigen' actions of the reference individual and the actions of its opponents.

When the system reaches an equilibrium state, i.e. the fluctuation of the action of each individual is equilibrated with the dissipation of the system, the probability that the action ${\bf \sigma^{i}}$ is part of the class of elements $C$ is given by
\begin{equation}
\rho^{i}=\frac{\sum_{c}\rho^{i}_{c}e^{-\beta \mathcal{H}^{i}}}{Z^{i}},
\label{DENSITY_AC}
\end{equation}
where $Z_{0}$ is the partition function on the storage distribution of the individual $i$ given by
\begin{equation}
Z^{i}=\sum_{c}e^{-\beta \mathcal{H}^{i}}.
\end{equation}
The dynamics of the actions depends on the size of the storage, which simultaneously determines the size of the internal energy associated to each individual. In the present context, it represents the change of the storage size of the individual $i$.

\subsection{Evolution of storage capacity}
The evolutionary dynamics is represented by a mutation process of the individuals. Each individual in the lattice is characterized by its storage size $m_{i}$. If the fitness is not large enough, the storage size changes; this change in the memory size automatically implies a change in the identity of the individual. The fundamental supposition is that the mutation of the memory size can increase the fitness of the individual. This mutation takes place only some times in a random way.

Such a process can be described by a Moran process (see Traulsen et. al.\cite{Traulsen}): (i) {\it selection} an individual is selected for reproduction with a probability related to its fitness; (ii) {\it reproduction} the individual produces one offspring; (iii) the offspring replaces a randomly selected individual. The microscopic state $l_{i}(t)$ represents the probability for a reproduction in a given time $t$, and can be either 1 if a reproduction takes place or 0 if not. Hence, the storage size of $i$ is given by $m_{i}(t+1)=m(t)+l_{i}(t)$. The transition probability of $l_{i}(t)$ is given by
\begin{equation}
V_{i}^{+}(t)=\Theta(f^{i}(t) - \sum_{j=1}^{k}f^{j}(t))\chi^{i}(t),
\end{equation} 
where $\Theta(x)$ is a heaviside step function, $f^{i}$ is the utility given by eq. (\ref{utility}) and $k$ is the number of neighbour individuals. The probability $\chi^{i}(t)$ is the probability that the individual $i$ has this mutation process. The dynamics of $l^{i}$ is given by
\begin{equation}
l^{i}(t+1) = l^{i}(t) + V_{i}^{+}(t)l^{i}(t),
\label{Master_II}
\end{equation}
this is the must simplest model of growth. Hence, equation (\ref{Master_I}) and (\ref{Master_II}) are two coupled equations describing the population's dynamics. The first describes how the information is handled by each of the individuals in the lattice; the second describes how the internal characteristics of the individuals evolves (storage capacity) depending on the individual's fitness. Both equations depend on the action each individual adopts. The first equation eventually reaches an equilibrium state after infinite times. Therefore, the single dynamical equation that remains is the equation for the mutation of memories. However, the fluctuation on equilibrium of the first equation produces a stochastic drive for the second one.
\subsection{Simulation method}
The system is defined on a square lattice. In each position of the lattice there is an individual doted with a storage size $m_{i}$ interacting with its four nearest individuals. Each individual is in a cell at the position $(X,Y)$ of the lattice. The whole system has periodic boundary conditions. Two Lattice sizes are considered in this work, with 4600 and 8100 individuals.
\begin{figure}[h!]
\begin{center}
\includegraphics[clip,  width=0.40\textwidth]{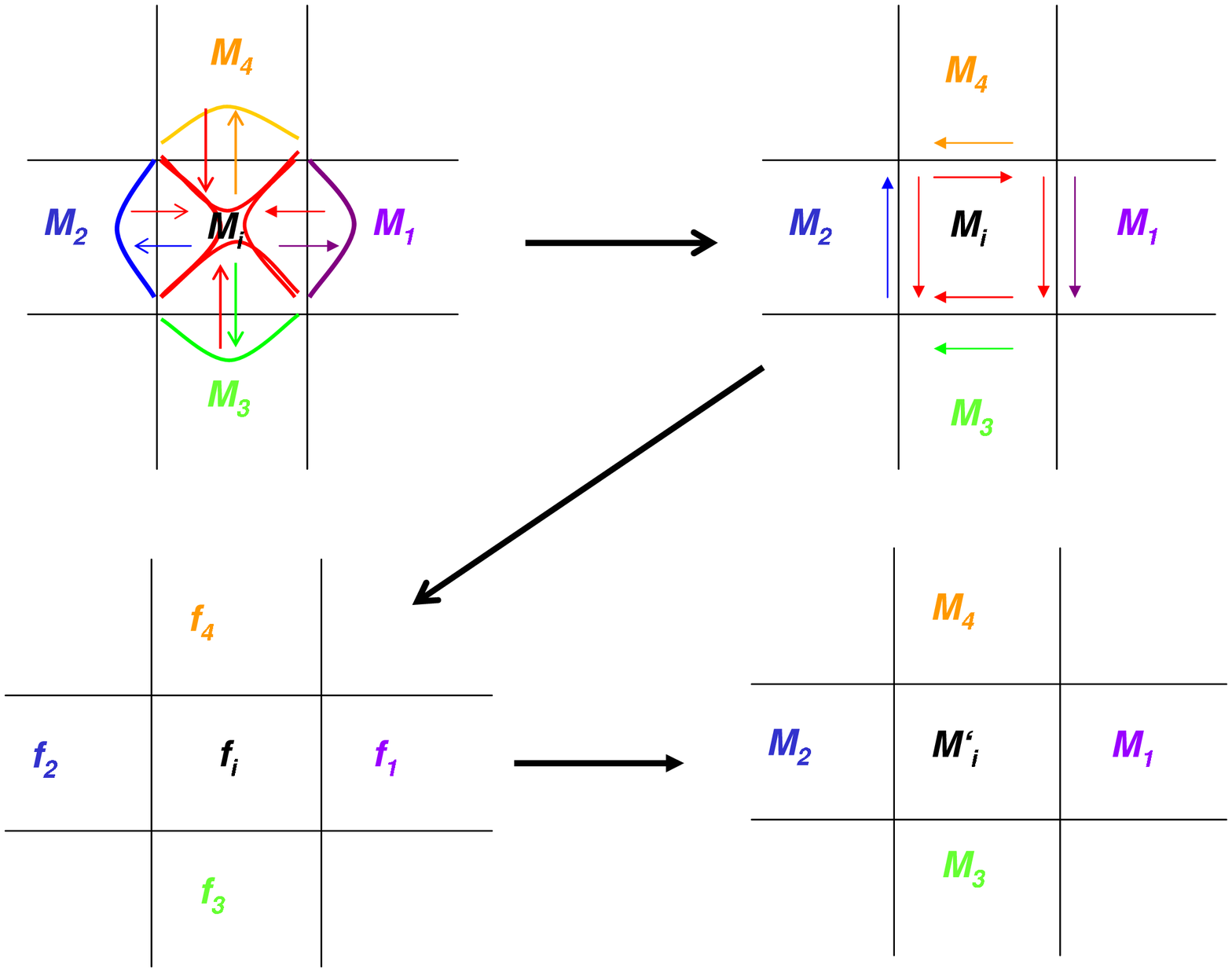}
\caption{Schematic representation of the simulation steps: perception and information processing, evaluation of new actions, evaluation of individual utilities and mutation of $m_{i}$ if individual's fitness is not large enough. This last step only takes place in a random way.}
\label{schema}
\end{center}
\end{figure}
The system is based on a Von Neumann Cellular Automata, i.e. each individual interacts with its four nearest neighbors. As the system is defined on pair interactions, each individual plays parallel games with the neighbors. For each action the individual uses a particular memory. This implies, each individual adopts different game strategies for different opponents. In each time unit the following process takes place (See Fig. \ref{schema}): each individual store the last own action and the action of its opponent in its memory (in the figure represented as a semicircle); with this information the individual decides its new action. In the next step the score of the game is computed; this information is the individual utility $f_{i}$ in each lattice position. If this utility is lower than the utility on the neighbors, and the individual is allowed to do a mutation, then a change of the memory size from $m_{i}$ to $m'_{i}$ takes place. This process is repeated in the next time step.

The system is initialized with random initial interactions and random initial memories. However, we want to observe how the evolution in the system takes place. Therefore, the first random memories in the system are small ones. Given that the system reaches an equilibrium state in the actions, the simulations are stopped until this state is reached. 

The dynamics of the system depends on the evolution of the memory size, as well as the learning parameters. In a real model such parameters should be defined in an individual way. However, for simplicity we assume that learning parameters are common for the whole individuals, such that it can be used as a control parameter.
\begin{figure}[h!]
\begin{center}
\includegraphics[clip,  width=0.40\textwidth]{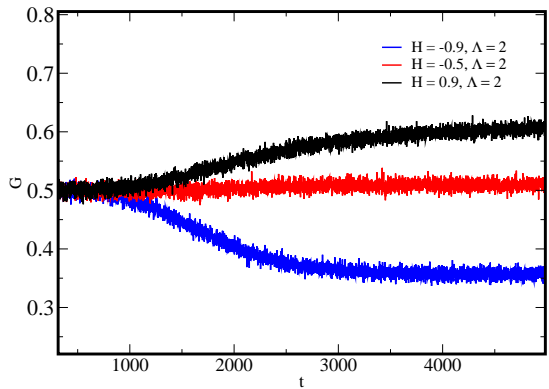}
\caption{Number of cooperators $\rho_{m}$ as a function of the time. In the inset the utility function is plotted as a function of the time. The dependence on the external field $H$ is shown.}
\label{fig_COOP_T}
\end{center}
\end{figure}

\section{Results}
\subsection{Distribution of actions}
The actions are discrete states for $C$ and $D$. After the system relaxes, the distribution of the number of each one of these states depends on the learning parameters. This dynamics is shown in Fig. (\ref{fig_COOP_T}), where the number of $C$ is computed for different external fields $H$. The distribution function of elements of class $C$ is given by
\begin{equation}
G[H,\Lambda](t) = \frac {\sum_{i=1}^{N} \rho^{i}(t)}{N},
\end{equation}
where $\rho^{i}$ is the probability that the individual $i$ is in the state $C$, defined by eq. (\ref{Master_I}), and $N$ is the total number of elements into the cell $X \times Y$. The number of cooperators can be tuned by means of the external field. For the first time steps there is an equilibrium state in the number of cooperators. But for $t>1000$ a 'split' in the function of number of cooperators takes place, depending on the external field. For $H>0$ the number of cooperators is larger than for $H<0$. This is equivalent to say that a set of individuals react to a cooperative attitude if the external influence is also cooperative; other wise, an external influence tending to defect may convince the individuals also to act as defect. This is equivalent to say that the individuals collectively adhere to a norm which is in this case imposed from outside.

\begin{figure}[h!]
\begin{center}
\includegraphics[clip, angle = -90,  width=0.30\textwidth]{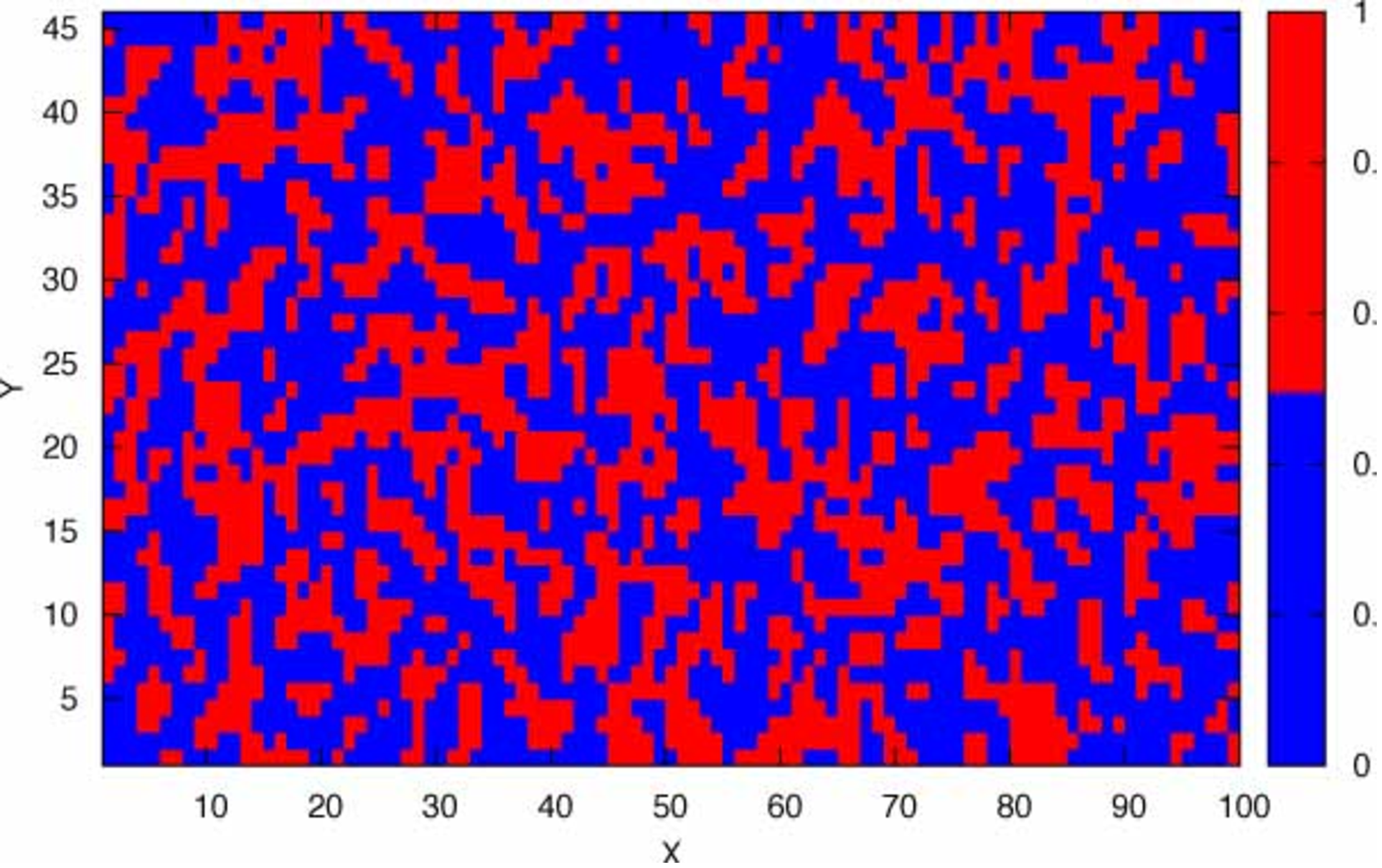}
\includegraphics[clip, angle = -90,width=0.30\textwidth]{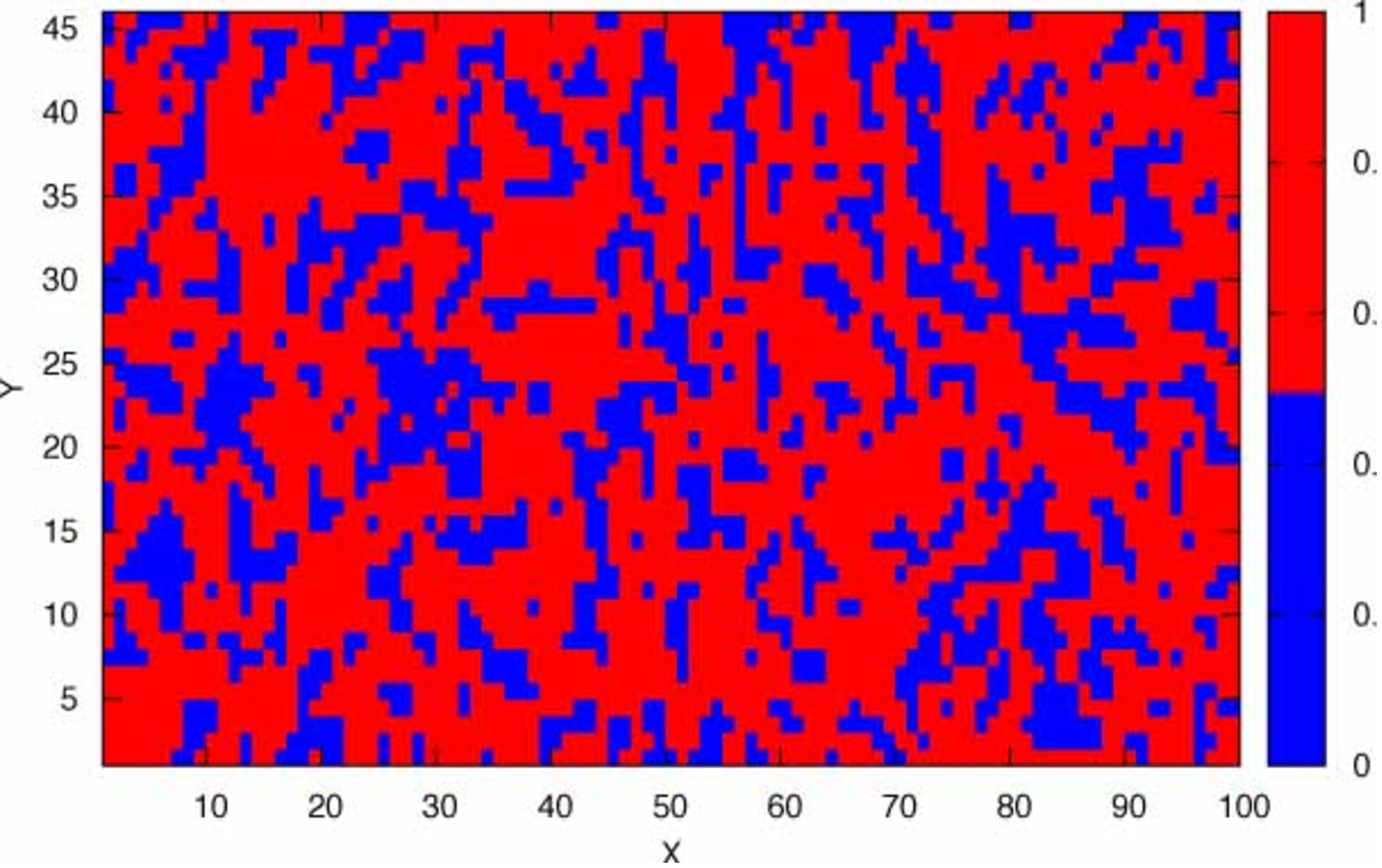}
\includegraphics[clip, angle = -90,  width=0.30\textwidth]{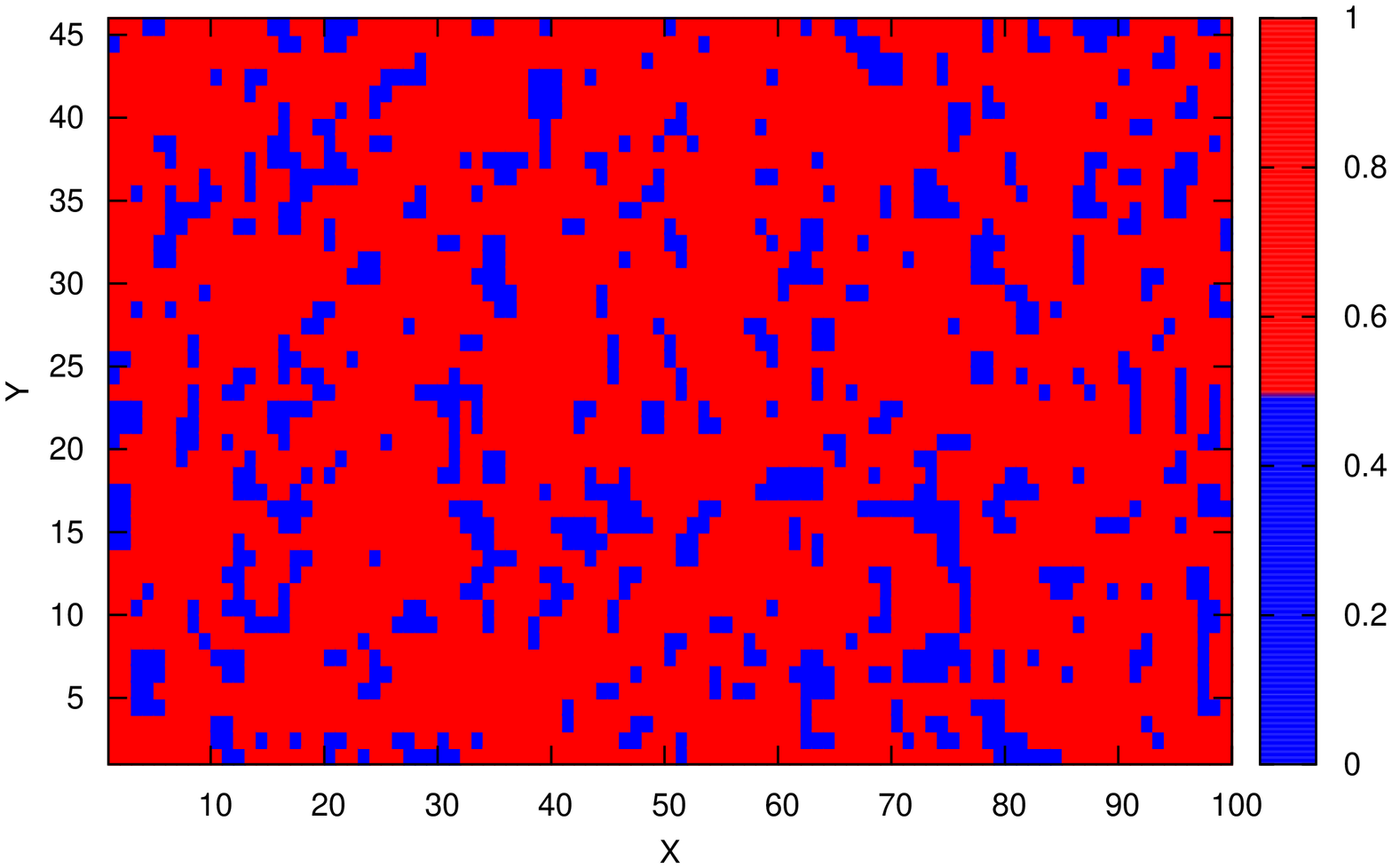}
\caption{Distribution of number of cooperators (in red) for equilibrium distributions in cells $X \times Y$ of size 46 $\times$ 100 elements.  First plot (left) is for $H=-0.9$, second (middle) for $H = 0.6$ and the last (right) is for $H = 0.9$. In the three cases, $\Lambda = 2$.}
\label{fig_Distr_C}
\end{center}
\end{figure}
The distribution of $C$ and $D$ into the lattice is shown in Fig. (\ref{fig_Distr_C}). This plot is done for a fix learning parameter and different external fields. The system shows a correlation between different individuals. However, no ordered domains are evident. For $H = -0.9$ there are large clusters of $D$. By tuning the field, the clusters of $D$ leaves place for clusters of $C$. The distribution of the clusters of cooperators for a positive external field should be symmetric to the distribution of defectors in a negative field. However, the snapshots show qualitatively a different cluster sizes of $D$ and $C$ at both extreme values of $H$. Hence, the distribution of either of both classes of individuals probably depends on the game, in particular on the fact that defect is more rewarded than cooperate.  
\begin{figure}[h!]
\begin{center}
\includegraphics[clip,  width=0.40\textwidth]{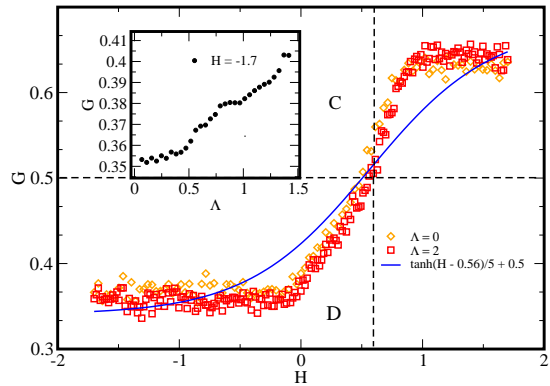}
\caption{Number of cooperators as a function of the external field $H$ for fixed $\Lambda$. In the inset plot the number of cooperators is drawn as a function of the learning parameter $\Lambda$ for $H<0$ fixed. For $H=0$ there is no dependence on the learning parameter.}
\label{fig_COOP_H}
\end{center}
\end{figure}
The dependence of the number of cooperators on the external field is shown in Fig. (\ref{fig_COOP_H}). The behavior of the system with respect to the external field is mathematically related to the behavior of a 1D Ising model. In such case the distribution of individuals $G$ is equivalent to the {\it magnetization} of the system. In the following analysis we denote $G[H]$ as a magnetization curve should be given by a hyperbolic tangent. Each individual plays a parallel game against its neighbors. Therefore, this function should scale proportional to the number of individuals playing different games. Since that the behavior of the number of cooperators is similar to the magnetization in an Ising model, the function that we use to fit this plot is defined as $G[H]=\frac{\tanh (H - H_{0})}{G_{0}} + G_{N}$, where $H_{0}$ is the shift on the field, $G_{0}$ is a scaling factor according to the scale of $G_{H}$ and $G_{N}$ is a scaling factor on $G$. If the memory size is constant, this dependence should be exactly equal to the fit function. However, the variation of the memory size, and the definition of the game in a square lattice, affects the smoothness of the magnetization curve. Therefore, the dynamics of the game has influence on the equilibrium distribution of cooperators and defectors and its dependence on external learning parameters.  


An interesting aspect is the dependence on the learning parameter $\Lambda$. Given that the learning schema is one dimensional, there is no possibility to observe a critical behavior as a function of $\Lambda$. However the dependence on this parameter is only observed for $H <0$, i.e. for a population under strong influence to defect an increase in the number of cooperators is related to the increase of $\Lambda$.

\begin{figure}[h!]
\begin{center}
\includegraphics[clip, angle = -90,  width=0.30\textwidth]{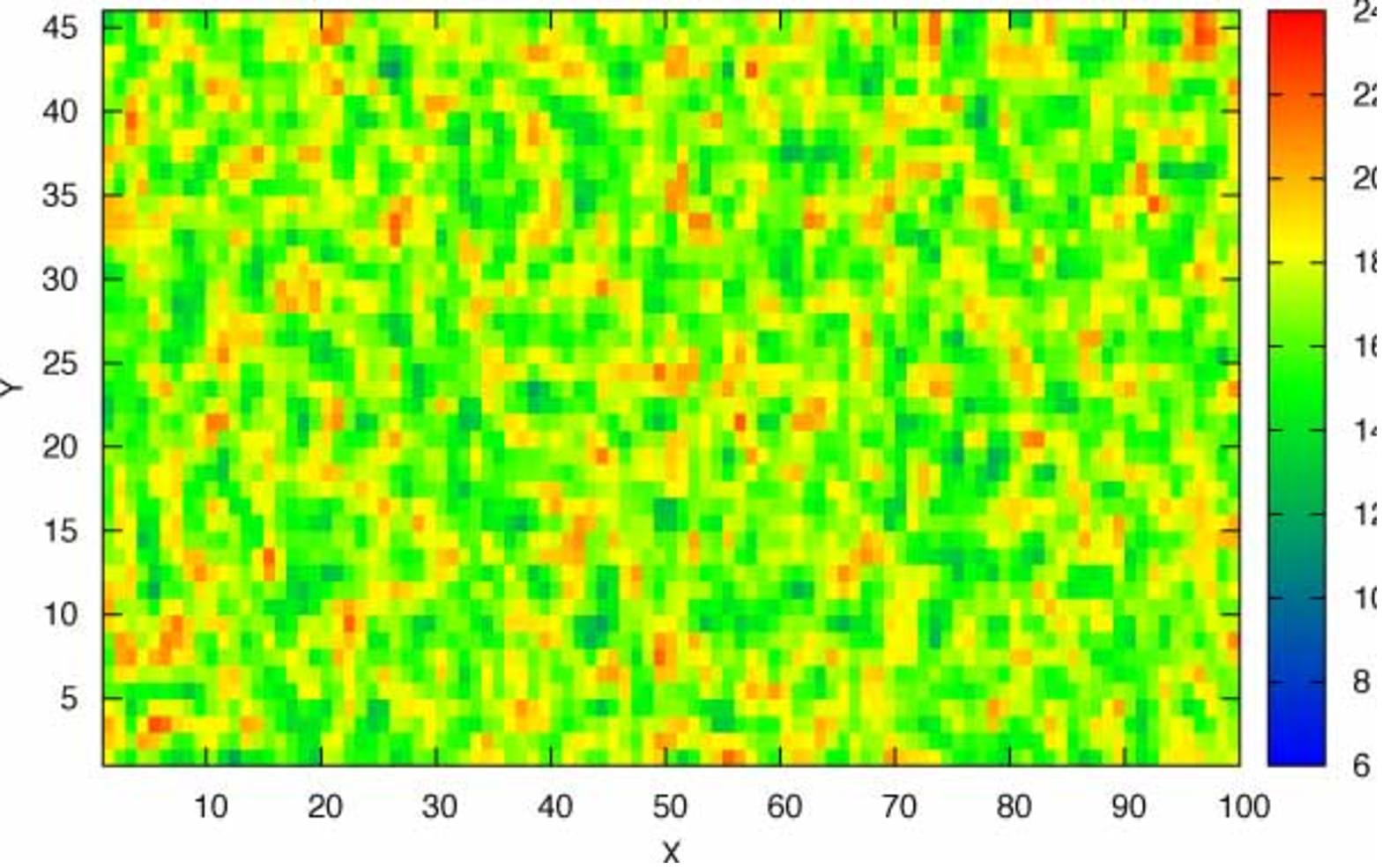}
\includegraphics[clip, angle = -90,  width=0.30\textwidth]{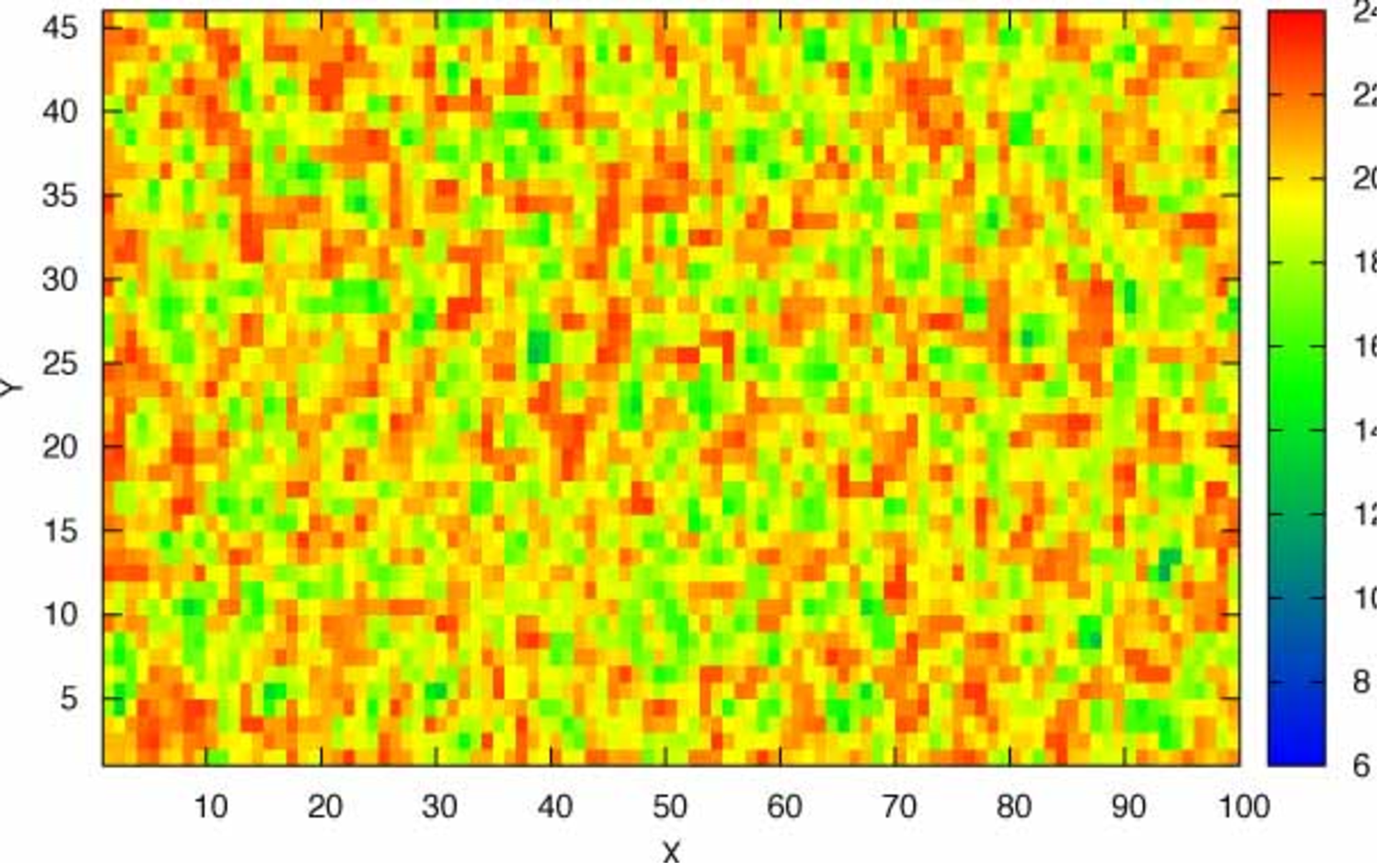}
\caption{Snapshots for the distribution of memories in a square lattice $X \times Y$ of size 46 $\times$ 100 elements. The first plot (left) is for $H=-0.9$; the second one (right) for $H = 0.9$. All plots are defined for a constant learning parameter $\Lambda = 2$. The colors represent the memory size in the Lattice.}
\label{fig_DISTR_M}
\end{center}
\end{figure}
\subsection{Distribution of memories} 
The memory is an important individual feature that may improve its response to evolution pressure. This distribution is defined as
\begin{equation}
\rho_{m}(t) = \frac{\sum_{i=1}^{N}m^{i}(t)}{N},
\end{equation}
where $N$ is the total number of elements into the system (please note that $\rho^{i}(t)$ denotes the distribution of cooperators, whereas $\rho_{m}(t)$ denotes the distribution of memories). This density is off course different to the density of cooperators $\rho(t)$.

A snapshot shows a distribution of memories in the square Lattice $X \times Y$. The colours represent the variation of different memory sizes, from 1 to 20, the maximal storage capacity defined for the present computations. The storage size distribution also strongly depends on the external field of the system. For $H < 0$ there is a relative uniform distribution of individual memories. By increasing the field, large memories are dominant. The rich structure for $H=0.9$ in fig. (\ref{fig_DISTR_M}) indicates a rapid evolution of large memories. This happens when the external field induces the system to be cooperative. A less rich but stable distribution for $H=-0.9$ is related to a more diversity, i.e. to the coexistence of different individuals with different storage sizes.

\begin{figure}[h!]
\begin{center}
\includegraphics[clip,  width=0.30\textwidth]{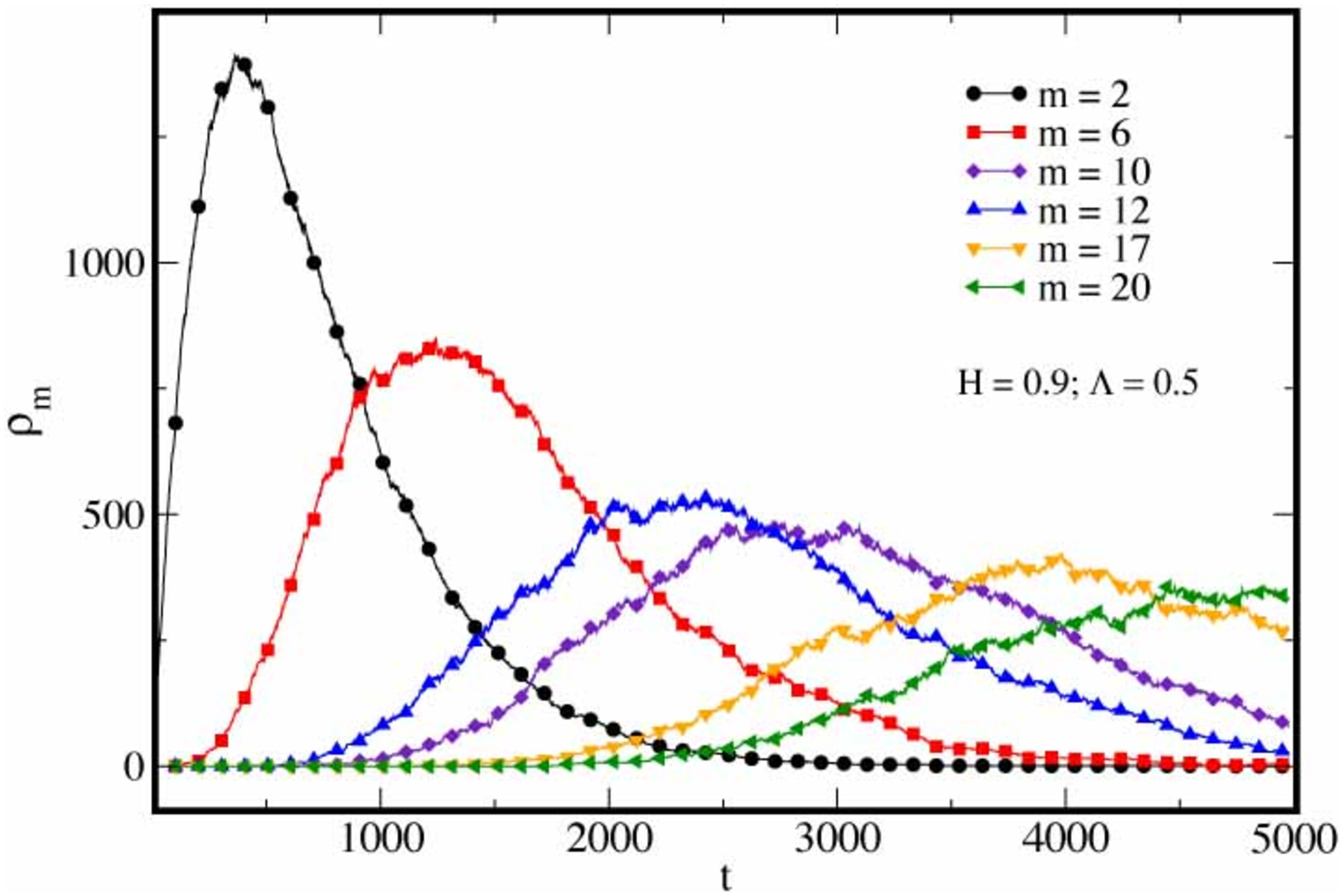}
\includegraphics[clip,  width=0.30\textwidth]{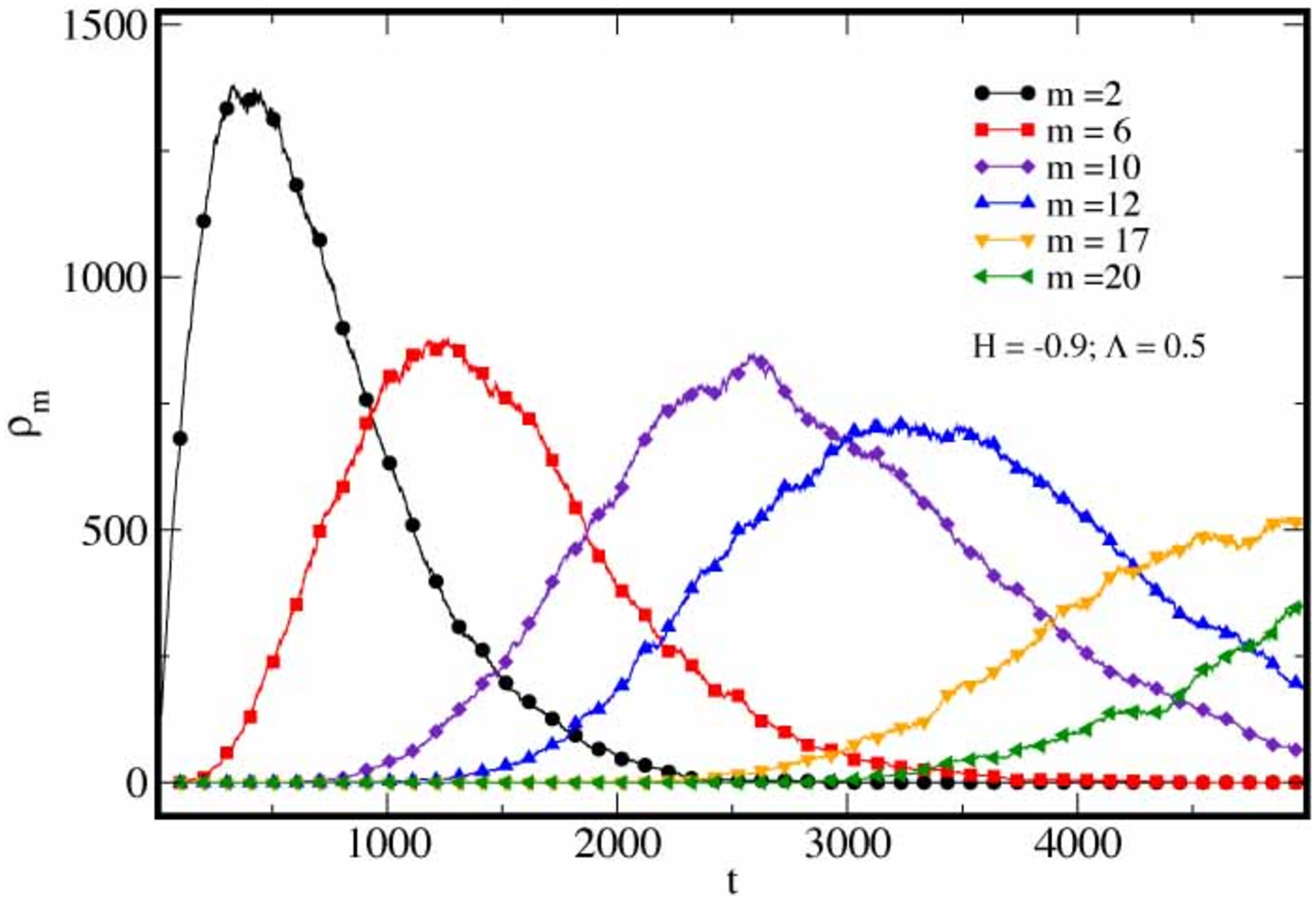}
\caption{Distribution of some memories as a function of time. The variation of the external field influences the diversity in the distribution of memories. For this reason, the distribution of memories is computed for two different external fields: $H=-0.9$ (left) and $H=0.9$ (right).}
\label{fig_M_DISTR_t}
\end{center}
\end{figure}
Here, the structure of the game is also related to the clusters with different size. The gradient in the memory size is relative small. Therefore, a phase separation between memories of different sizes takes place. This effect is similar to the formation of domains composed by individuals with similar memory. 

The complexity and diversity of distribution of memories can be analyzed in the time. In Fig. (\ref{fig_M_DISTR_t}) the distribution of number of individuals with a memory $m_{i}$, for few memories, as a function of the time is shown. The distribution in the time is an asymmetric function. Each individual $m_{i}$ has a growing rate, related to a high fitness, and a subsequent decay rate, related to the decrease of the fitness because new individuals appear into the population. At the same time individuals with less memory sizes face extinction from the system. Eventually, this individual with memory $m_{i}$ faces also extinction when the fitness function is not large enough. The relaxation process is related to the growth of a new individual with a different memory size. This result shows the rivality between individuals and the dependence of the memory size on the fitness function of the system. In the extinction process the distribution functions have a long tail, while the memory $m$ still preserves a high fitness.
\begin{figure}[h!]
\begin{center}
\includegraphics[clip,  width=0.40\textwidth]{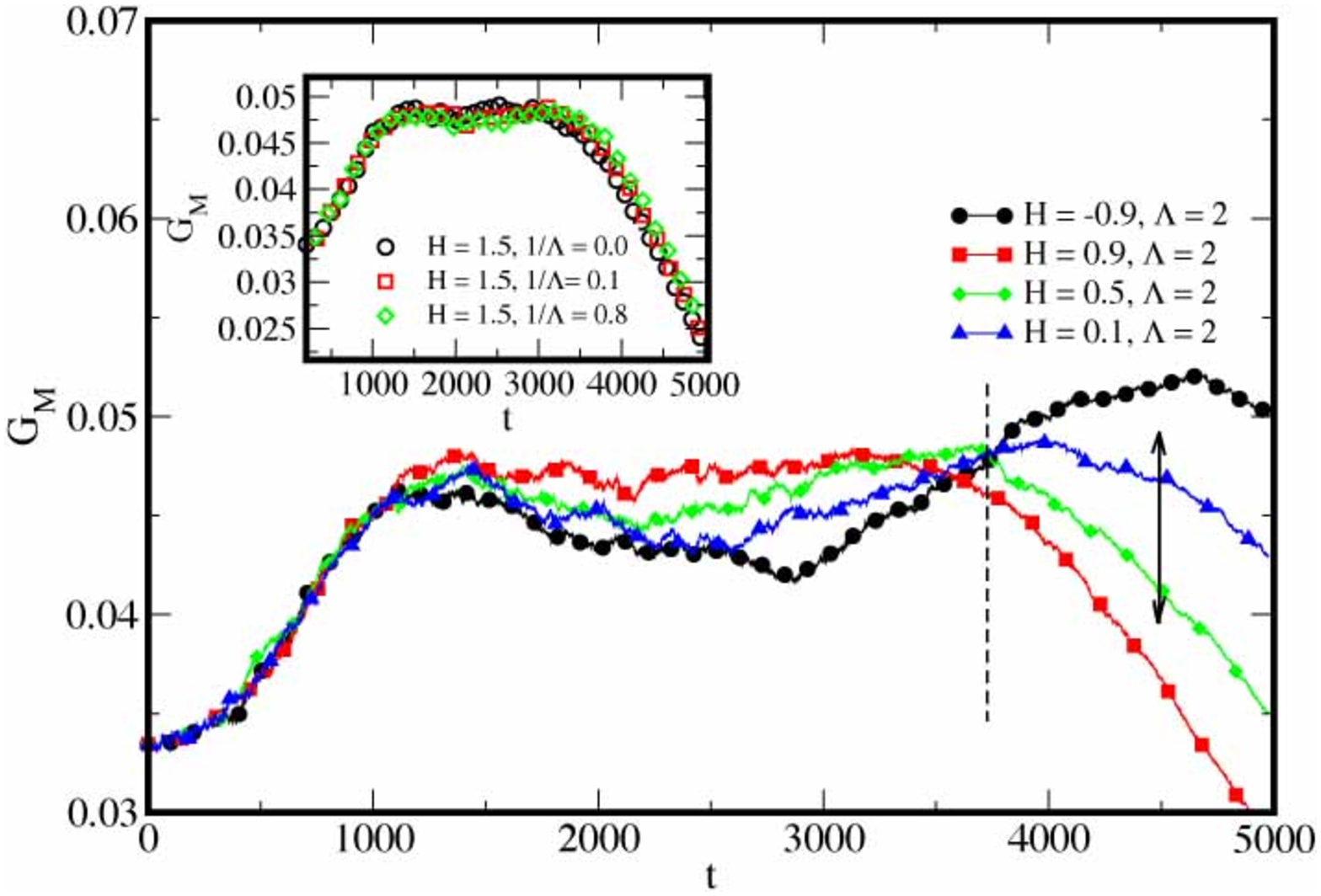}
\includegraphics[clip,  width=0.40\textwidth]{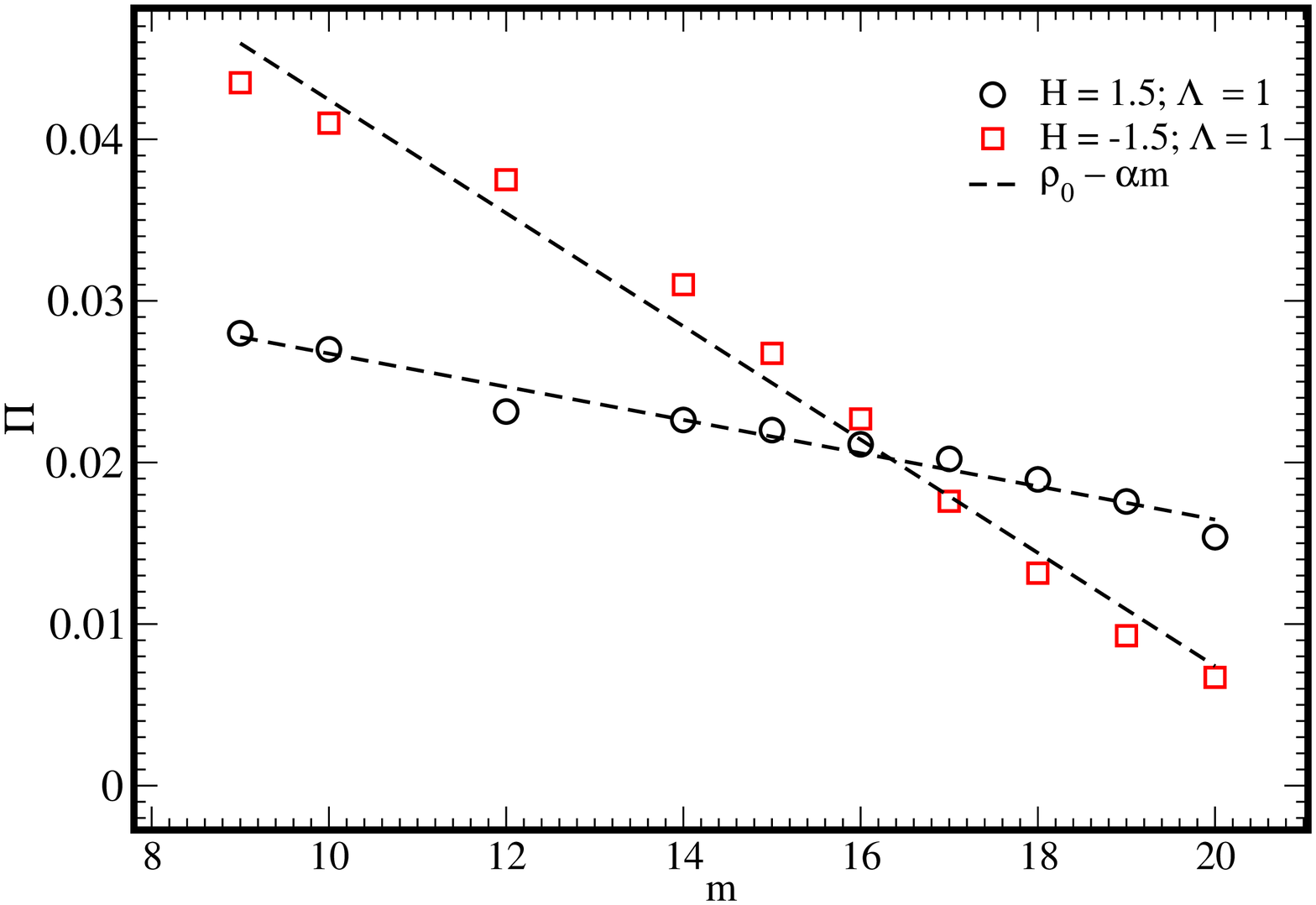}
\caption{Average number of the total number of memories $G_{M}(t)$ as a function of time (left plot). In the second plot (right) we show the total density of the number of individuals $\Pi(m)$ as a function of the memory size. Here, the slope of $\Pi(m)$ depends on the external field $H$.}
\label{fig_M_t}
\end{center}
\end{figure}
For two external fields the distribution of memories in the time shows a change in the distributions. As is shown in Fig. (\ref{fig_M_DISTR_t}), for $H=-0.9$ (Defect) the tails of the distributions are fat in comparison to $H=0.9$ (Cooperate). Hence, the extinction rate can be controlled using the external field.

In general the distribution $G_{M}(t)$ of memories in the time shows how the effect of the change of $H$ is, where
\begin{equation} 
G_{M}[H,\Lambda](t)=\frac{\sum_{m} \rho_{m}(t)}{N},
\end{equation}
 $N_{M}$ the number of memories defined into the system. For short and intermediary times there is no dependence on the field because the system is still plenty of relative small memories (Fig. \ref{fig_M_t}). At a critical time $t_{c}$ new individuals with larger memories co-evolve. After this time, i.e. for long time regimes, a dependence on the external field introduces a change into the tails of the curves. The variation as a function of the external field is more dramatic as the dependence on the learning parameter $\Lambda$ (See the inset of Fig. (\ref{fig_M_t})). It is also interesting to note that the total density of $m_{i}$, defined as 
\begin{equation}
\Pi[m]=\frac{\int_{o}^{\infty} \rho_{m}(t)dt}{N \times T},
\end{equation}
 is proportional to the memory size, i.e $\Pi = \Pi_{0} - \alpha m$. In every case the slope depends on $H$ and $\Lambda$. According to the result that we obtain, $\alpha_{1}>\alpha_{2}$ if $H_{1}>H_{2}$. 

The number of individuals in a population is proportional to the storage size. In a extinction process of the population, the different decays in the number of individuals of each species depends on the size of the window where this process takes place. Hence, the size of the extinction channel is equivalent to the memory (the size of the genetic code) of each individual.

A very important additional aspect related to the distribution of densities $\rho$ is the measurement of the diversity of the system. There are several measures for diversity (See for example the Simpson's index \cite{Simpson}). We adopt a Shannon's index, which provides information not only about the population richness, but also the composition of the community. This index is given by \cite{Kleidon, Krebs} 
\begin{equation}
I = -\sum_{m=1}^{N_{M}} \Pi[m] \log \Pi[m],
\label{Shannon}
\end{equation}  

where $N_{M}$ is the number of memories. The Shannon's equitability, a measure of diversity quantifying how equal the population are numerically, can be calculated by dividing the index $I$ by the maximal Shannon's index $I_{Max}$ defined by $D_{Max} = \log N_{M}$
\begin{equation}
I_{E} = \frac{I}{I_{Max}}.
\end{equation}

The diversity $I$ is therefore a function of $\alpha$ and $\rho_{0}$; in particular the diversity is a function of the information introduced into the system. For the present work $I = 0.43$ for $H = -0.9$, whereas $I=0.39$ for $H=0.9$, implying that an external influence for defect increases the diversity of the system. Therefore, the diversity can be tuned by means of the external field. This is a very important result, because it implies that diversity is not only related to the amount of energy but also is related to the amount and kind of information it is introduced into the system. In both cases, the evenness $I_{E}$ is equal to the diversity, i.e. for $H = -0.9$ the evenness is larger than for $H=0.9$. The relation between diversity and information flow into the system is interesting. If this prediction is right, then the diversity cannot only be argument on energy flux but also on information process by the elements of the system.

\begin{figure}[h!]
\begin{center}
\includegraphics[clip,  width=0.40\textwidth]{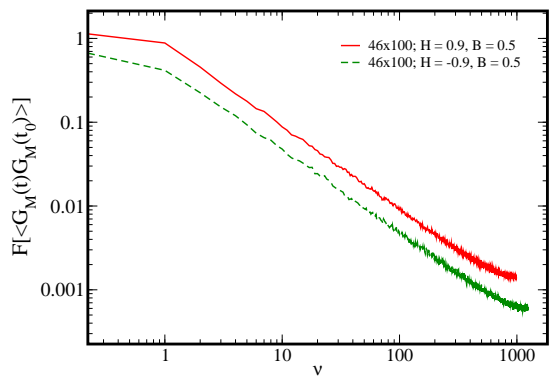}
\caption{Power spectrum for the distribution of memories. The curves fit with a function of the form $\sim \nu^{-1}$, indicating a critical state in the evolution of memories.}
\label{T_fig_M_t}
\end{center}
\end{figure}


The evolution of memories is a non equilibrium process. This fact is corroborated in Fig. (\ref{fig_DISTR_M}), where different domains are distributed in the $X \times Y$ lattice. This is a process with different scales that may also appear at different time scales. In this work we use the function $G_{M}(t)$ as an input signal in order to compute the power spectra of the distribution of memories.

The power spectra of  $G_{M}(t)$ (See Fig. (\ref{T_fig_M_t})) can be fitted with a curve of the form $\sim \nu^{-\alpha}$, where $\nu = 2\pi / t$. The results obtained in the present investigation fit with $\alpha \sim 1$. This implies that there are long correlations in the time in the aggregate number of memories. In particular, given the definition of $<N[M]>$ is possible to conclude that this spectra arises from the superposition of individual evolutionary processes. Therefore, the co-evolutive process is in this aspect equivalent a system with self organized criticality. However, the power spectra does not depend on the learning parameters.

\section{Discussion} 
The obtained results should be analysed from two points of view. {\it First}, the present model reproduces basic characteristics of the Ising learning in the lattice of players. However the variation of the memory size, related to the fitness of each individual, introduces variations in the actions of the individuals. The coupling of the system to the co-evolutive dynamics does not introduce phase transitions, but this dynamics, together with the dimension of the lattice where the simulations were performed, affect the dependence on the learning parameters. 

{\it Second}, the dynamics of the game influences the memory distribution into the system. The variation of the external field influence the complexity of the distribution of memories and, therefore, the diversity of the distribution of individuals characterized by $m_{i}$.

Therefore, an important conclusion is that the {\it learning characteristics} of the individuals may influence the distribution of $m_{i}$. Although the co-evolutive process is far from equilibrium, it may be influenced by external information imprinted into the system. The dimensionality of the storage device of the individuals is an important factor. A one dimensional storage device does not allows that $\Lambda$ may introduce a change in the phase behavior of the actions of the individuals. Like more dimensional Ising models, increasing of the dimension may introduce interesting effects that may have influence in the phase behavior of the distribution of $m_{i}$. 

The final relevant result in this investigation is the {\it role of the external information} has in the {\it entropy} of the system. Using this model we are able to analyse the genetic diversity of a population, which is the neg-entropy of the system. An external information source, changing the phenotype of the population, has an influence on the diversity of the system. In particular, the diversity increases if a large part of the population has a normative tendency to defect. This fact implies a reduction of the entropy of the system. In other words, {\it diversity} is not only a function of the size of the population \cite{Andayani}, but also {\it depends on cognitive characteristics and external sources of information}. This result can be a good approach to explain, why in some regions there are more diversity than in their neighbors, taking into account that genetic diversity is correlated to the community structure \cite{Crutsinger}. A possible answer in such case is that animals, water or wind transport genetic material or signals that may influence the information exchange of the individuals. This phenomenon is not only exclusive in biology. Intuitively, it is equivalent, for instance, to the effect of background music and advesting on consumption behavior of individuals in shopping malls. In this last case, the external information, and not only budget, is the essential factor that affects individual behavior \cite{Lesourne}.

The model implemented is a very simple approach that joints fundamental aspects from the theory of ferromagnetism with population dynamics in the frame of learning theory. This model can be extended to different directions. From a physical point of view it is interesting to make an analysis of the dimensionality of the memory in the individual's behavior. For high dimensional memories is possible to expect a critical behavior in the distribution of cooperators. In biology is interesting to make an analysis of the effect that an inhomogeneous and fluctuating external information source may have in the distribution of the population. Finally, an extension from the lattice to a different topology in the interaction of agents may have interesting effects in the population's distribution.

Several scholars have in the last time discussed, if natural selection is a causal process at the population's level or if it takes place at individual's level. By accepting frequency dependent selection this causality is placed at the population level. However, if the individuals express a kind of cognition, related to the reaction to noise and information sources, then this causality must also account for the individual. In resume, in the present model we propose middle point into this dispute. By tuning external information as well as noise sources we are able to observe how same individuals may affect the population. By observing a mutation process we observe how the population influences the distribution of individuals. We have shown that this coupling have an important connection in the measure of diversity of the population. In particular, we showed that avoiding a cooperative behavior is possible to increase this diversity.



\end{document}